\documentstyle[12pt,aaspp4]{article}

\begin{document}

\title{The Metallicity Dependence of the Stellar Luminosity and Initial
Mass Functions: HST Observations of Open and Globular Clusters}

\author{Ted von Hippel\altaffilmark{2,3}}
\affil{Department of Astronomy, University of Wisconsin, Madison, WI 53706, USA}
\authoremail{ted@noao.edu}

\author{Gerard Gilmore, Nial Tanvir, David Robinson, 
and Derek H.P. Jones\altaffilmark{4}}
\affil{Institute of Astronomy, Madingley Road, Cambridge CB3 0HA}

\altaffiltext{1}{Based on observations with the NASA/ESA {\it Hubble Space
Telescope}, obtained at the Space Telescope Science Institute, which is
operated by AURA, Inc, under NASA contract NAS 5-26555.}

\altaffiltext{2}{Institute of Astronomy, Madingley Road, Cambridge CB3 0HA}
\altaffiltext{3}{address: WIYN Telescope, NOAO, PO Box 26732, Tucson, AZ 
85726-6732, USA}
\altaffiltext{4}{Royal Greenwich Observatory, Madingley Road, Cambridge CB3 0EZ}

\begin{abstract}

\bigskip

Using HST and the WFPC2 we have acquired very deep V- and I-band
photometry of stars in NGC 2420 and NGC 2477 to study cluster luminosity
functions at approximately solar metallicity.  We have determined these
cluster luminosity functions down to $M_I$ = 10.5 (0.2 M$_{\odot}$) and
find that the luminosity function of NGC 2420 turns over at $M_I$
$\approx$ 9.0, and possibly stops altogether by $M_I$ $\approx$ 9.5.  The
luminosity function of NGC 2477 may flatten at $M_I$ $\geq$ 9.5.  We
compare our open cluster luminosity functions to the solar neighborhood
field star luminosity function of Kroupa, Tout \& Gilmore (1993) and the
four published HST globular cluster luminosity functions:  $\omega$ Cen
(Elson {\it et al.}\ 1995), 47 Tuc (De Marchi \& Paresce 1995b), M 15 (De
Marchi \& Paresce 1995a), and NGC 6397 (Paresce, De Marchi \& Romaniello
1995).  We find a smooth relation between the location of the luminosity
function turn-over and the metallicity for all these low mass star samples
which matches the expected $M_I$ versus [Fe/H] trend for a model star of
$\approx$ 0.27 M$_{\odot}$ (Saumon 1995; Alexander {\it et al.}\ 1996).
We interpret this smooth and systematic behavior in the cluster luminosity
functions as strong evidence in favor of an invariant initial mass
function and a metallicity-dependent mass-luminosity relation.

\end{abstract}


\section{Introduction}

One of the most fundamental aspects of the stellar content of galaxies is
the distribution by mass of newly formed stars, the initial mass function
(IMF).  Evidence supporting or restricting variations in the IMF for very
low mass stars is of profound importance for an understanding of the mass
content of the universe, the dynamical evolution of open and globular
clusters, the chemical evolution of galaxies, and the physics of star
formation.  While developments in star formation theory (e.g.  Shu 1991)
and observations (e.g. Evans 1991) continue to increase our understanding
of the complexities and processes involved in star formation, any
understanding of how the IMF does, or does not, depend on environment and
metallicity is still elusive.  For low mass stars, which are the subject
of this paper, large surveys (see Reid \& Gilmore 1982; Hawkins \& Bessell
1988; Stobie, Ishida \& Peacock 1989; Bessell \& Stringfellow 1993) have
provided a fairly well determined measure of the field star luminosity
function near the Sun.  Extensive modelling is required to derive the
corresponding stellar initial mass function.  However, there is reasonable
agreement (cf. Kroupa {\it et al.}\ 1993 and Tinney 1995 for recent
references) that the luminosity function shows a broad maximum near $M_I
\approx +9.5$, while the underlying mass function (MF) breaks away from an
approximate power-law increase below 0.5 M$_{\odot}$.  Star count surveys
in the near-IR (e.g.  Hu {\it et al.}\ 1994) and in the optical with HST
(Bahcall {\it et al.}\ 1994; Santiago, Gilmore \& Elson 1995), and most
recently from the Hubble Deep Field (Elson, Santiago \& Gilmore 1996),
suggest that the field stars of the Galactic thick disk and halo have a
luminosity function which is not significantly distinguishable from that
of the solar neighborhood down to masses near the hydrogen burning limit.
This is a remarkable result, given that these stellar populations differ
considerably in local density, and by a factor of forty in mean chemical
abundance.

Field star analyses are, however, subject to many uncertainties, in
particular imprecise distance determination (Kroupa {\it et al.}\ 1993).
The derivation of an initial mass from a color and apparent magnitude of
an object which may be an unresolved binary star of unknown age,
distance, and metallicity remains a statistical goal, rather than a
precise tool.  By contrast, in clusters the primary uncertainties of
differential ages, abundances and distances are removed.  Hence, extensive
surveys of the open clusters near the Sun have been undertaken (e.g.
Pleiades: Hambly \& Jameson 1991; Hyades: Reid 1993; Praesepe: Williams,
Rieke \& Stauffer 1995; Hambly {\it et al.}\ 1995).  Such clusters are
however primarily young, so that interpretation suffers from the
limitation that evolutionary models for pre-main sequence stars of low
mass are not yet well tested.  Ideally, one wants clusters older than
perhaps 1 Gyr, covering a wide range of metallicities.

In the past year there has been a vast improvement in the quality and
quantity of data on the low mass end of the stellar luminosity function
(LF) from observations of globular clusters with the refurbished HST
(e.g.\ Elson {\it et al.}\ 1995; De Marchi \& Paresce 1995a, 1995b;
Paresce, De Marchi \& Romaniello 1995).  These four globular cluster
studies have already provided good stellar LFs to $M_I \geq 10$ ($\approx$
0.2 M$_{\odot}$), and they will soon be supplemented by at least seven
more HST globular cluster studies.  In this paper we report deep HST
observations of the LFs in two open clusters, which  allow us to extend
the metallicity range represented by these four globular clusters ($-2.2 <
[Fe/H] < -0.7$) up to solar abundance.  Thus, for the first time we now
have cluster luminosity functions extending to very low stellar masses and
covering more than 2 dex in metallicity, and can therefore study the
effect of metallicity on stellar luminosity functions.  In this paper we
search for a systematic dependence of the IMF on some parameter to provide
clues to the physics of star formation.

\section{Observations and Data Reduction}

We observed the open clusters NGC 2477 and NGC 2420 with HST on 1994 March
18 and 1994 May 18/19, respectively, using the WFPC2 along with the F555W
(approximately V-band) and F814W (approximately I-band) filters.  Our
observations were taken 5.2 arc minutes SW of the cluster center
(7:38:11.2 +21:33:09.7, J2000.0) for NGC 2420 and 2.7 arc minutes WSW of
the cluster center (7:52:16.8 -38:35:40.1, J2000.0) for NGC 2477 to avoid
the brightest stars in each cluster.  These fields still contain many very
bright stars so we obtained the cumulative one hour exposures in each
filter from nine 400s exposures in NGC 2477 and four 900s exposures in NGC
2420.  All data were acquired in the same manner and with the same
filters, although the operating temperature of the CCDs was $-76^{\circ}$C
during the March observations and $-88^{\circ}$C during the May
observations.  The temperature change was made in order to decrease a
charge-transfer efficiency (CTE) problem which manifested itself as a
position-dependent non-linearity of the WFPC2 CCDs.  The effect was found
to be $\approx$ 10\% at $-76^{\circ}$C and $\approx$ 4\% at $-88^{\circ}$C
(Holtzman {\it et al.}\ 1995b) in a series of 40 second calibration
exposures taken of a field in $\omega$ Cen.  Elevated sky counts found in
longer exposures such as ours are expected to decrease the CTE problem
(Holtzman {\it et al.}\ 1995a).  The lower operating temperature also
resulted in far fewer `hot pixels'; pixels with an elevated dark count
lasting for hours or days.  Finally, an additional difference between our
two cluster observations is that the sky level for the May data was five
times higher than for the March data, owing to zodiacal light
differences.

The raw data were reprocessed using the IRAF/STSDAS software with the
updated calibration frames that became available subsequent to the
observations.  This included the standard statistical corrections to the
analogue to digital conversion, bias subtraction, dark count subtraction,
flat fielding and geometric distortion correction.  We also included the
non-standard ``ramp correction'' for the March CTE problem.  The CTE
properties of the WFPC2 are still not entirely understood, but applying
the correction reduced the significance of the CTE effect to $\approx$
1\%.  The CTE effect in the May data is expected to be $\leq$ 1\% because
of the higher sky counts and the cooler CCDs, so we did not attempt to
apply a correction for these data.  Additionally we spent great effort
trying to identify hot pixels as measured in the dark frames taken shortly
before or after our observations, but were largely unsuccessful owing to
the short life-times of these hot pixels.  Since hot pixels have counts
which are independent of the filter in use, and thus have color equal to
zero on the instrumental system, they appear in the same region of the
color-magnitude (CM) diagram as the faintest white dwarfs.  Since the
WFPC2 CCDs are undersampled, hot pixels may appear like faint stars,
especially two adjacent hot pixels or single hot pixels next to positive
sky fluctuations.  Hot pixels thus decreased our limiting magnitude for
white dwarfs (von Hippel {\it et al.}\ 1995, hereafter Paper I), but did
not affect the limiting magnitude for the coolest main sequence stars, the
subject of this paper.

We performed aperture photometry using 1.5 pixel radius apertures with
aperture corrections based on a detailed study of the many bright stars
available in our fields.  After some experimenting we did not employ point
spread function (PSF) fitting photometry as our fields were uncrowded and
the PSF was positionally dependent and not well characterized by our
stars.  Unlike Paper I, the photometry was directly transformed to the
standard Johnson-Cousins VI system using the calibration from Holtzman
{\it et al.}\ (1995b).  The photometry we present here is thus slightly
different from that presented in Paper I, but the differences are so
small for the white dwarfs as not to modify the results of that study.
The CM diagrams for all the stars in the fields of both our clusters, with
galaxies removed, are presented in Figures 1 and 2.

\subsection{Completeness Estimates}

In order to characterize the completeness of our photometry we created
synthetic stars using the TinyTim HST PSF generation package (Krist 1994),
added these stars to our data, and then tried to recover them using our
standard finding and photometry procedure.  The synthetic stars were
created for both the F555W and F814W filters and were centered at 16
different sub-pixel positions and laid down in a grid of positions
covering the whole of each CCD.  This sub-pixel sampling was done since
the PSF locations affect the number of pixels under the PSF, and thus the
sky noise and read noise contributions per pixel cause the limiting
magnitude to depend on the PSF sub-pixel position.  Since this effect is
based on noise changing systematically with PSF position, it affects
completeness, but not the measured magnitudes of objects bright enough to
find.  The resulting simulated noise-free stars were scaled and had noise
added to them to mimic stars over a range of magnitudes.  From this
procedure we estimated our completeness levels, which are superimposed on
Figures 1 and 2.  These completeness levels lead to a limiting magnitude
$\approx$ 0.5 magnitudes brighter than initially expected for the
refurbished HST, but this is now understood as due to the CCDs leaking
some charge between adjacent pixels (Holtzman {\it et al.}\ 1995a).  These
estimates of completeness should be reasonably accurate for the redder
stars where hot pixels are not a problem.

\subsection{Uncorrected effects and remaining sources of uncertainty}

We outline here the remaining sources of uncertainty in our photometry,
although we emphasize that these are all small effects which do not affect
our results.

1) The CTE corrections are almost surely not perfect for the
   $-76^{\circ}$C data and they have not been attempted for the
   $-88^{\circ}$C data.  The resulting uncertainties are $\approx$ 1\% and
   do not contribute to any important systematics.

2) The HST pipeline flats were taken at the gain=14 setting, whereas our
   data were taken at the gain=7 setting.  The relative gains between
   these two settings are not identically 2.0 for all CCDs.  The
   appropriate gain ratios have been applied, but these are uncertain to
   $\approx$ 1\% each.  This source of uncertainty should not cause
   systematic errors since all three WF CCDs contain nearly equal portions
   of the cluster CM diagram.

3) The HST pipeline flats are good to a few percent, pixel-to-pixel, over
   most of the area of the CCDs.

4) We applied a single aperture correction to all photometry for all WF
   CCDs.  In fact the relatively large number of bright but isolated stars
   in our images allowed us to investigate a number of subtle instrumental
   effects in the photometry.  We performed a number of such tests, and
   compared them with simulations based on TinyTim analytical PSFs.  The
   results are described in detail in Tanvir, Robinson \& von Hippel
   (1995) but briefly they found that (a) the photometry is essentially
   unaffected by the position of the center of the star within a pixel,
   and (b) there is a small ($\approx \pm$3\%) variation seen in the NGC
   2477 small aperture photometry which correlates with position on the
   CCD.  This latter effect is probably due to small variations in the PSF
   across each CCD.  A similar but not identical effect is seen in the
   simulations, while NGC 2420 has too few bright stars to show it.
   However, given that this effect is small, is uncorrelated with
   magnitude, and to first order affects V and I in the same way, thus
   leaving V-I colors unchanged, it makes a negligible difference to the
   results presented here and so no correction was made.

5) The precision of the Holtzman {\it et al.}\ (1995b) transformation is
   expected to be $\approx$ 2\%, over the applicable color range.  The
   range of applicability of the Holtzman {\it et al.}\ photometric
   transformations is $-0.3 <$ V-I $< 1.5$, whereas our stars continue to
   V-I $\approx$ 3.5.  The expected error for the reddest stars is not
   large (see Section 3), but this is probably our single greatest
   systematic uncertainty.

6) Whitmore (1995) reports that the HST + WFPC2 throughput increased
   by 4.1\% in F555W and by 0.6\% in F814W after cooling the WFPC2 CCDs to
   $-88^{\circ}$C on 24 April 1994.  Holtzman {\it et al.}\ (1995b) report
   similar numbers.  Since the photometric zero points were determined
   after the CCDs were cooled, corrections need to be applied to the
   earlier data, in our case to NGC 2477.  Most of the throughput
   differences seem to be due to the decrease in the CTE problem
   (Ferguson, private communication), however, and since we make a CTE
   correction for our NGC 2477 data, much of this zero point difference
   probably disappears.  The zero point difference between CTE corrected
   data has not yet been determined.  We therefore have not applied any
   throughput correction to the NGC 2477 data.  Thus, our NGC 2477 data
   may suffer an additional $\approx$ 2-5\% systematic error in V-I.

 From items 1 through 5 above we estimate that the additional random error
contributed by the instrument, above and beyond that produced by photon
counting statistics, is likely to be 4-5\%, largely independent of
magnitude.  Item number 5 is relevant to the color magnitude diagrams
presented in Figures 1a, b and 2a, b as well as to any comparison of these
data in the transformed Johnson-Cousins system to models.  It does not
affect our analysis of cluster luminosity functions as we compare our data
to other HST observations taken through the same filters.  Also, these
additional noise sources are not included in the CM diagram error bars,
which represent only the internal photon-counting errors and read-noise.

\section{Analysis}

In this section we present the CM diagrams for NGC 2420 and NGC 2477 and
derive their main sequence LFs.  NGC 2420 is metal-poor for an open
cluster, with $[Fe/H] \approx -0.45$, is only lightly reddened, with a
uniform E$\rm _{B-V} = 0.05$, and has a distance modulus 11.95.  Ages from
isochrone fitting for this cluster range from 2.1 Gyrs (Carraro \& Chiosi
1994) to $3.4\pm0.6$ Gyrs (Anthony Twarog {\it et al.}\ 1990) and 4 Gyrs
(VandenBerg 1985).  We derived (Paper I) a lower limit for the age of this
cluster from the white dwarfs luminosities of 1.3 Gyrs.  NGC 2477 is more
metal rich, with $[Fe/H] \approx 0$, is heavily and unevenly reddened,
with an adopted E$\rm _{B-V} =0.33$ (Smith \& Hesser 1983; Hartwick,
Hesser \& McClure 1972), and has a distance modulus of about 10.6.
Isochrone fitting ages for this cluster range from $0.6\pm0.1$ Gyrs
(Carraro \& Chiosi 1994) to $1.2\pm0.3$ Gyrs (Smith \& Hesser 1983).  We
derived (Paper I) a lower limit on the age of this cluster from the white
dwarfs luminosities of 1.3 Gyrs.

The photometry for NGC 2420 and NGC 2477 is presented in the CM diagrams
of Figures 1a and 1b, respectively.  The photometry has been transformed
to the Johnson-Cousins system and 1 $\sigma$ error bars are plotted for
all objects.  Completeness levels derived from our simulations are
indicated at the bottom of these figures by the dotted lines.  The
completeness levels are labeled and correspond to $\approx$ 90, 50, and
10\% completeness.  The V-band completeness levels are nearly horizontal,
being curved slightly due to the color term in the photometric
transformations.  Likewise the I-band completeness levels are at a
near-constant value in I, and thus inclined in this CM diagram.  The
saturation limits are indicated by the upper dotted lines, level for V and
inclined for I.  The reddening vector for each cluster is plotted at V
$\approx$ 22 and V-I $\approx$ 0.6.  Examination of these figures shows
that the main sequence stars and white dwarfs clearly stand out in both
clusters, even in the relatively crowded field of NGC 2477 (which has
galactic latitude $b$ = $-5.8$).  It is also evident that the completeness
simulations are consistent with the photometric errors and the faint limit
of the photometry.

Figures 2a and 2b present the cluster photometry after the effects of
distance and reddening have been removed, and after the removal of the
obvious contaminating field stars.  The dashed lines are eye-ball fits to
the Monet {\it et al.}\ (1992) trigonometric data for white dwarfs and
solar metallicity main sequence stars.  They are plotted here to
demonstrate that the entire combination of extrapolated photometric
transformations, measured distances, and reddenings are reasonable for
these clusters.  Figure 2b presents the fits to the Monet {\it et al.}\
data placed for E(B-V) = 0.2, 0.3 and 0.4 (left to right), since this
cluster is known to have variable reddening.  Contaminating field stars
were removed by comparison with nearby fields observed by the HST Medium
Deep Survey project.  The Medium Deep Survey CM diagrams (Santiago,
private communication) were derived from F606W and F814W photometry, so we
used the transformations derived by Elson {\it et al.}\ (1995) to convert
to the Johnson-Cousins colors.  A detailed comparison was then made of the
number of stars as a function of magnitude and color in one nearby field
for NGC 2420 and two comparable fields for NGC 2477.  The comparison field
for NGC 2420 is at {\it l,b} = 206, +19.6, whereas NGC 2420 is at {\it
l,b} = 198, +19.7.  The comparison field has a similar, and nearly
negligible, reddening, E(B-V) = 0.02 (Burstein \& Heiles 1982).  Two
comparison fields were used for NGC 2477 since none yet exist which
closely match its Galactic position and reddening.  One field is at {\it
l,b} = 85, -3.5 and has E(B-V) $\geq$ 0.39, whereas the other is at {\it
l,b} = 202, +9 and has E(B-V) $\geq$ 0.18 (Burstein \& Heiles 1982).  NGC
2477 is at {\it l,b} = 254, -5.8.  Using this procedure we found it
straightforward to isolate the main sequence cluster stars from the
contaminating foreground and background stars, most of which lie well away
from the main sequences.  A few stars in the comparison fields do lie on
the cluster main sequences, and thus a small proportion of the stars
presented in Figures 2a and 2b are expected not to be cluster members.
These can only be removed statistically, so no corrections were applied to
Figures 2a and 2b, though the statistical corrections were applied to the
cluster main sequence LFs presented below.  Before moving on to the
extracted cluster LFs, it is worth pointing out that the main sequence in
NGC 2420 (Figure 2a) may actually stop at $M_V$ $\approx$ 12 ($M_I$
$\approx$ 9.5), as the fainter stars lie somewhat off the expected main
sequence locus.  Since we argue below that this cluster LF peaks at $M_V$
$\approx$ 11.5 ($M_I$ $\approx$ 9.0), for the present purposes we make the
conservative assumption that these stars are cluster members.

The main sequences LFs for NGC 2420 and NGC 2477 are plotted in Figures 3a
and 3b.  The star symbols are the raw LFs, and the open circles are the
LFs corrected for field star contamination and incompleteness.  Also
plotted are lines representing the fractional incompleteness based on our
synthetic star simulations.  The fractional incompleteness rises from
$\approx$ 0\% at $M_I$ = 10 to nearly 100\% by $M_I$ = 11.5.  In all
subsequent use of the cluster LFs we include only portions of the LF where
incompleteness is less than 25\%.  The NGC 2420 LF (Figure 3a) turns over
at $M_I$ $\approx$ 9.0 while the NGC 2477 LF (Figure 3b) seems to flatten
at $M_I$ $\geq$ 9.5.  The numbers of stars in these LFs are small, being
only 19 and 53 stars in the complete portion of the LFs for NGC 2420 and
NGC 2477, respectively, so these LF shapes are currently only suggestive.
We can estimate the likelihood that the NGC 2420 LF does not turn over by
evaluating the probability that the value of the LF in the final bin is
actually greater than the identified peak.  For NGC 2420 a turn-over is a
1.5 $\sigma$ result, i.e.  there is a 7\% chance that this LF continues to
rise to its measured limit.  The likelihood that the LF continues to rise
as rapidly after $M_I$ $\approx$ 9.0 as it did before this point is only
1\%, however, since the last LF bin lies 2.3 $\sigma$ below such an
extrapolation.  For NGC 2477 a turn-over is not seen, and a flattening is
only suggestive.  We thus locate the peak in this LF at, or after, $M_I$ =
9.5.  We proceed below to consider the implications of these LF shapes,
assuming that they have been correctly determined.

In Figure 4 we plot our open cluster LFs along with the solar neighborhood
LF presented in Kroupa {\it et al.}\ (1993) and the four published HST
globular cluster LFs: $\omega$ Cen (Elson {\it et al.}\ 1995), 47 Tuc (De
Marchi \& Paresce 1995b), M 15 (De Marchi \& Paresce 1995a), and NGC 6397
(Paresce, De Marchi \& Romaniello 1995).  The four globular cluster LFs
have been shifted down by 1 to 2 units in the log to approximately match
the open cluster number counts, for ease of presentation.  Since all the
HST results were obtained using the F814W filter with WFPC2, we choose to
plot the LFs directly in terms of F814W magnitudes, and have therefore
transformed the Kroupa {\it et al.}\ (1993) data first to $M_I$ using the
trigonometric parallax data of Dahn {\it et al.}\ (1995), and then to
$M_{814}$.  Error bars are not plotted here for clarity, but the globular
cluster LFs are well determined with typically many hundreds of stars per
bin near the LF peak.  Error bars due to Poisson statistics are large for
our open cluster data and can be readily determined either in this figure
or in Figures 3a and 3b by reference to the vertical scale.

In Figure 5 we plot the location of the turn-over of all the LFs of Figure
4 with respect to their metallicity, measured in [Fe/H].  The location of
a LF turn-over is defined as the location of the peak bin, or for the case
of $\omega$ Cen, the location of the first bin of the LF flattening.  The
solar neighborhood value has been placed at [Fe/H] = $-$0.2,
representative of this sample (Wyse \& Gilmore 1996).  The abundance
distribution of the field stars (unlike the clusters which each have a
single abundance) blurs any sharp features in its LF, so that the LF
maxima discussed here are less precisely defined for the solar
neighborhood.  The error in the turn-over magnitude for the solar
neighborhood is $\approx$ 0.25 magnitudes, while the error in the
turn-over magnitude for NGC 2420 is larger, $\approx$ 0.5 magnitudes.  The
turn-over magnitude for NGC 2477 is plotted as a lower limit.  The
internal errors in the turn-over magnitudes for the globular clusters are
small because of the large number of stars in each bin.  The errors in the
turn-over magnitudes are thus likely dominated by the distance
uncertainties for these clusters, which are $\approx$ 0.2 magnitudes (e.g.
Elson {\it et al.}\ 1995).  We do not plot the location of the turn-over
in the very broad halo field star LF as given by Dahn {\it et al.}\ (1995)
in Figure 5, as the location of this peak is probably more uncertain than
the other LF peaks.  

The essential result which is evident in Figure 5 is a smooth and well
defined correlation between the location of the turnover of the luminosity
function of a stellar sample and the metallicity of that stellar sample.
While we discuss explanations of this correlation below, we note here that
it bears a close resemblance to a theoretical constant-mass locus.  Such
loci are plotted in Figure 5 as dashed lines from the low mass star models
of Saumon (private communication), for masses $0.2$ and $0.231$
M$_{\odot}$, and as solid lines from the low mass star models of Alexander
{\it et al.}\ (1996), for masses $0.2$, $0.25$ and $0.3$ M$_{\odot}$.  The
Saumon models are plotted in $M_I$, rather than $M_{814}$, but the
differences are expected to be minor.

\section{Discussion}

In the preceding sections we have concentrated on the cluster luminosity
functions, not the mass functions, since the derivation of the latter
requires additional assumptions.  Indeed, Elson {\it et al.}\ (1995), in
their analysis of $\omega$ Cen, show that even small uncertainties in the
mass-luminosity relation are enough to cause major changes in the derived
mass function slope.  D'Antona \& Mazzitelli (1996) also demonstrate that
the faint star LF slope has little to do with the MF slope, or
metallicity, as it is driven by the slope of the mass-luminosity
relation.  While theoretical models for low mass stars are rapidly
improving (see Alexander {\it et al.}\ 1996; D'Antona \& Mazzitelli 1996),
a number of difficulties still remain at solar metallicities.  We
therefore chose not to convert these LFs to MFs, but rather to follow the
dominant LF feature, the turn-over, and compare it with stellar models.
By following the LF turn-overs we are able to see that all available LFs
look essentially the same, especially when corrected for the metallicity
of the population.  All LFs have approximately the same shape, and each
has a turn-over at $\approx 0.27$ M$_{\odot}$.  Why might all LFs have
turn-overs at the same mass value, and what physical phenomena might
affect the turn-over?  We look at the effects of dynamics, the
mass-luminosity relation, and the IMF slope in turn.

In analyzing their globular cluster luminosity functions, De Marchi \&
Paresce (1995a, 1995b) argue that while these three globular clusters have
undergone different degrees of core collapse and tidal stripping, their
observed LFs are relatively insensitive to the cluster dynamical histories
since they were obtained near the cluster half-mass radii.  Our open
cluster observations were also made well inside the radii of tidal
stripping, 3 pc from the center of NGC 2420 and 1 pc from the center of
NGC 2477.  These distances should be compared to the tidal radius for NGC
2420, which was determined to be $\geq$ 18 pc by Leonard (1988).  The
tidal radius for NGC 2477 has not been determined to the best of our
knowledge, but these clusters are similar in their stellar density and
distance from the Galactic center.  Additionally, although NGC 2420 has a
relaxation time of only $\approx$ $10^8$ yrs, it shows only marginally
significant mass segregation (Leonard 1988).  Furthermore, since the
globular clusters have a mean stellar density a factor $10^3$ to $10^5$
times higher than the open clusters (compare the densities of Leonard 1988
with those of Webbink 1985), and are approximately an order of magnitude
older, it would require careful tuning of the dynamical histories of these
clusters for low mass stellar evaporation to produce, just at the present
time, an uncoordinated set of changes which mimic a systematic dependence
on metallicity.  In any case, several more globular clusters have been
observed with HST, so additional information to check the importance of
dynamical evolution, and the true scatter in the correlation of Figure 5,
will soon be available.  We do not argue that these clusters have not
undergone possibly very different dynamical histories, only that their
dynamical histories are not the primary cause of the LF turn-overs.

The slope of the mass-luminosity (ML) relation, in both V and I, is
rapidly increasing at masses lower than $\approx 0.27$ M$_{\odot}$,
independent of metallicity (Alexander {\it et al.}\ 1996; D'Antona \&
Mazzitelli 1996).  Metallicity does cause an offset in luminosity at a
given mass, but it only slightly affects the slope of the ML relation at a
given mass (Alexander {\it et al.}\ 1996).  This increase of slope in the
ML relation causes a given mass increment to cover a larger and larger
absolute magnitude increment, especially below M$_I$ $\approx$ 9.5 - 10.
Thus a turn-over in the LF at these magnitudes is naturally explained by
the rapidly steepening ML relation.  This point was appreciated by Kroupa
{\it et al.}\ (1993) for the solar neighborhood LF, and the theoretical
models of Alexander {\it et al.}\ (1996) and D'Antona \& Mazzitelli (1996)
now support this interpretation for all metallicities.  A test of this
hypothesis is whether these ML relations, when applied to our cluster LFs,
yield smooth MFs.  When we apply either the observed solar metallicity ML
relation of Henry \& McCarthy (1993) or the theoretical ML relation of
Alexander {\it et al.} (1996) to our LFs, we do indeed find MFs which rise
smoothly through the peak and to the completeness limit of the data.  This
statement cannot be made strongly, however, due to the small number of
stars in our LFs.  We note that our measured MFs rise as N $\sim$ M$^{-1}$
for both clusters, but that this value for the slope should not be taken
too seriously, for the reasons mentioned above with regard to the
insensitivity of this procedure to the MF slope, and because of its great
dependence on the slope of the ML relation.  In fact, D'Antona \&
Mazzitelli (1996) demonstrate that a change in the IMF slope moves the
peak of the LF to a different mass value, by approximately 1 magnitude in
the I-band for a change of 0.5 in the MF slope, depending somewhat on
metallicity.  The LF peak is thus a sensitive measure of the MF slope, and
any significant changes in this slope from cluster to cluster would have
destroyed the match between the data and the single mass locus seen in
Figure 5.

The increase in the slope of the ML relation below M$_I$ $\approx$ 9.5
makes a smooth mass function a viable explanation for these cluster LFs.
It is therefore unnecessary to invoke the third possibility, a MF kink at
$\approx 0.27$ M$_{\odot}$, as the cause of the LF turn-overs.  Such a MF
kink is still possible, since the MF is very poorly determined at lower
masses, but it is not {\it necessary}.

A complication we have not addressed here is the effect on the actual MF
of undiscovered binaries.  A large number of stars with masses below
$\approx 0.27$ M$_{\odot}$ could be hiding as unseen companions of
brighter main sequence stars.  Furthermore, globular clusters, open
clusters, and the solar neighborhood field could have different relative
numbers of such low mass companions.  A detailed analysis of the effect of
binarism on luminosity functions was presented by Kroupa {\it et
al.}\ (1993) to which the reader is referred.

We conclude that the low mass IMF is, to first order, independent of
metallicity, with all the systematic changes in luminosity functions with
metallicity dominated by the systematic changes in the mass-luminosity
relation with metallicity.  The implication for star formation is that a
large range of initial conditions and environments produce essentially the
same distribution by mass of low mass stars.  Additionally, the low mass
MF is not steep enough to provide a large number of essentially invisible
main sequence stars (Kroupa {\it et al.}\ 1993).

\section{Conclusions}

Using HST and the WFPC2 we have acquired very deep V- and I-band
photometry of stars in NGC 2420 and NGC 2477 to study cluster luminosity
functions at approximately solar metallicity.  We have determined these
cluster luminosity functions down to $M_I$ = 10.5 (0.2 M$_{\odot}$) and
find that the LF of NGC 2420 turns over at $M_I$ $\approx$ 9.0, and
possibly stops altogether by $M_I$ $\approx$ 9.5.  The LF of NGC 2477 may
flatten at $M_I$ $\geq$ 9.5.  We compare our open cluster luminosity
functions to the solar neighborhood field star luminosity function of
Kroupa {\it et al.}\ (1993) and the four published HST globular cluster
luminosity functions:  $\omega$ Cen (Elson {\it et al.}\ 1995), 47 Tuc (De
Marchi \& Paresce 1995b), M 15 (De Marchi \& Paresce 1995a), and NGC 6397
(Paresce, De Marchi \& Romaniello 1995).  We find a smooth relation
between the location of the luminosity function turn-over and the
metallicity for all these low mass star samples which matches the expected
$M_I$ versus [Fe/H] trend for a model star of $\approx$ 0.27 M$_{\odot}$
(Saumon 1995; Alexander {\it et al.}\ 1996).  We interpret this smooth and
systematic behavior in the cluster luminosity functions as strong evidence
in favor of an invariant initial mass function and a metallicity-dependent
mass-luminosity relation.

\acknowledgments
We thank Becky Elson, Basilio Santiago, Geraint Lewis and the HST User
Support Branch for advice and assistance and Didier Saumon for calculating
stellar models for us.  We also thank an anonymous referee for her/his
careful reading which led to a much improved presentation.  TvH was
supported by a PPARC fellowship during this work.

\vfill\eject

\figcaption[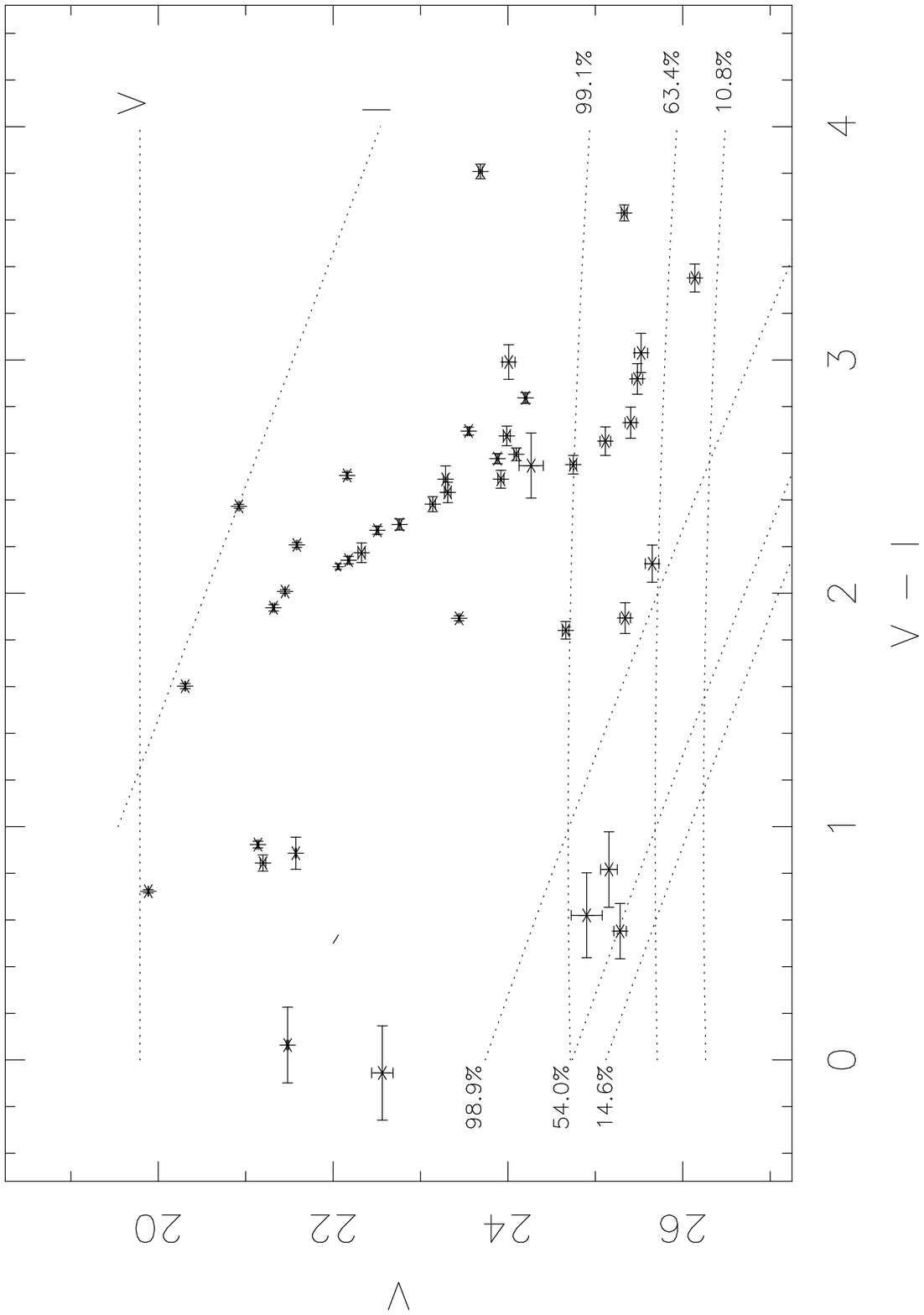,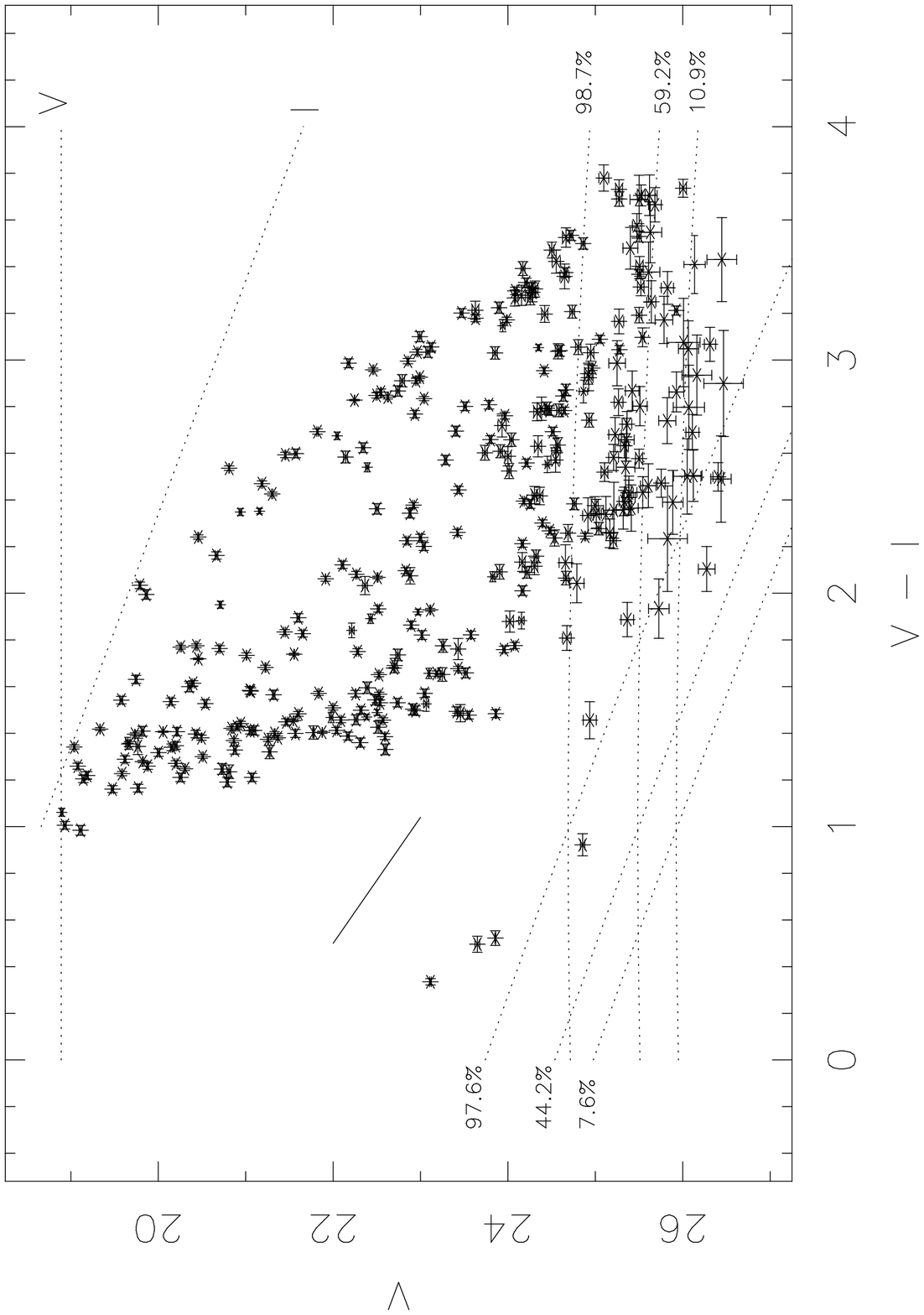]
{Color-magnitude diagrams for (a) NGC 2420 and (b) NGC 2477.  The error
bars represent one sigma internal errors in the photometry.  Photometric
saturation lines and labeled completeness levels for V and I are drawn at
the tops and bottoms of the figures, respectively.  Also plotted are the
reddening vectors for each cluster, which have not been applied.}

\figcaption[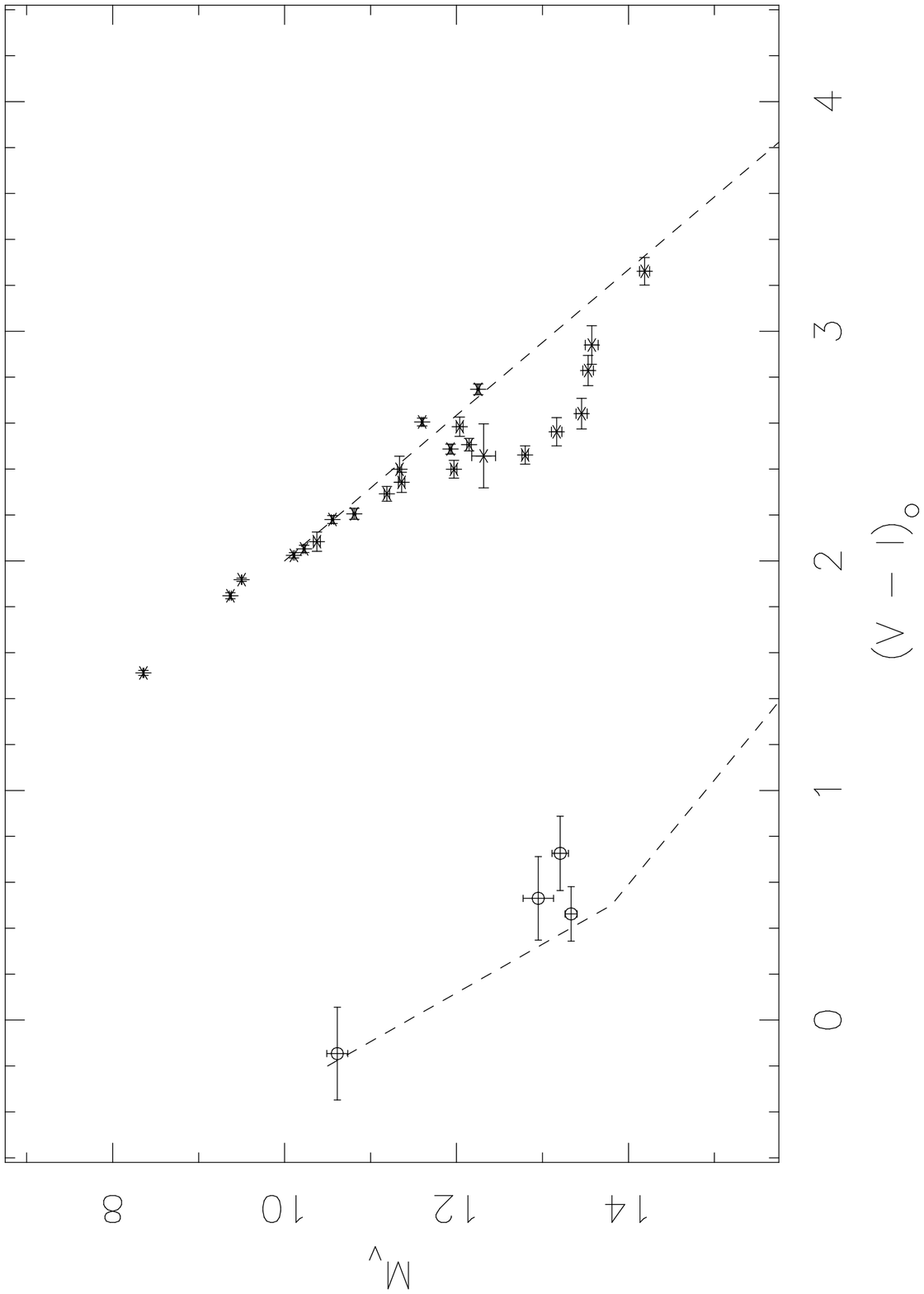,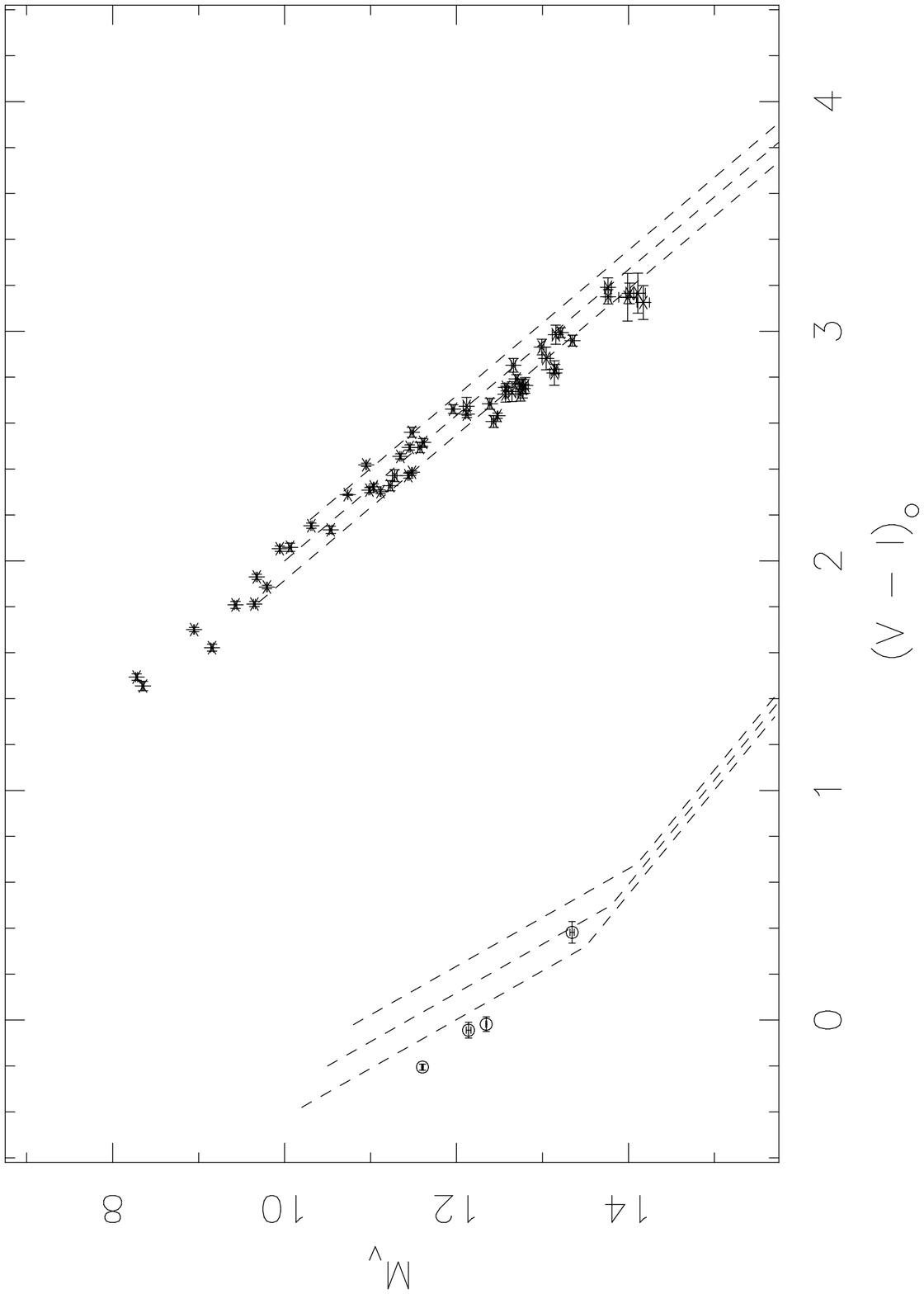]
{Similar to Figures 1a and 1b except background and foreground stars have
been removed, and the cluster has had its distance modulus and reddening
removed.  Also plotted in (a) NGC 2420 and (b) NGC 2477 as dashed lines
are the locations in the HR diagram for solar neighborhood stars from
Monet {\it et al.}\ (1992).  Figure 2b is plotted with 3 solar
neighborhood sequences representing, from left to right, E(B-V) = 0.2, 0.3
and 0.4 (canonical E(B-V) = 0.33), since this cluster is differentially
reddened.}

\figcaption[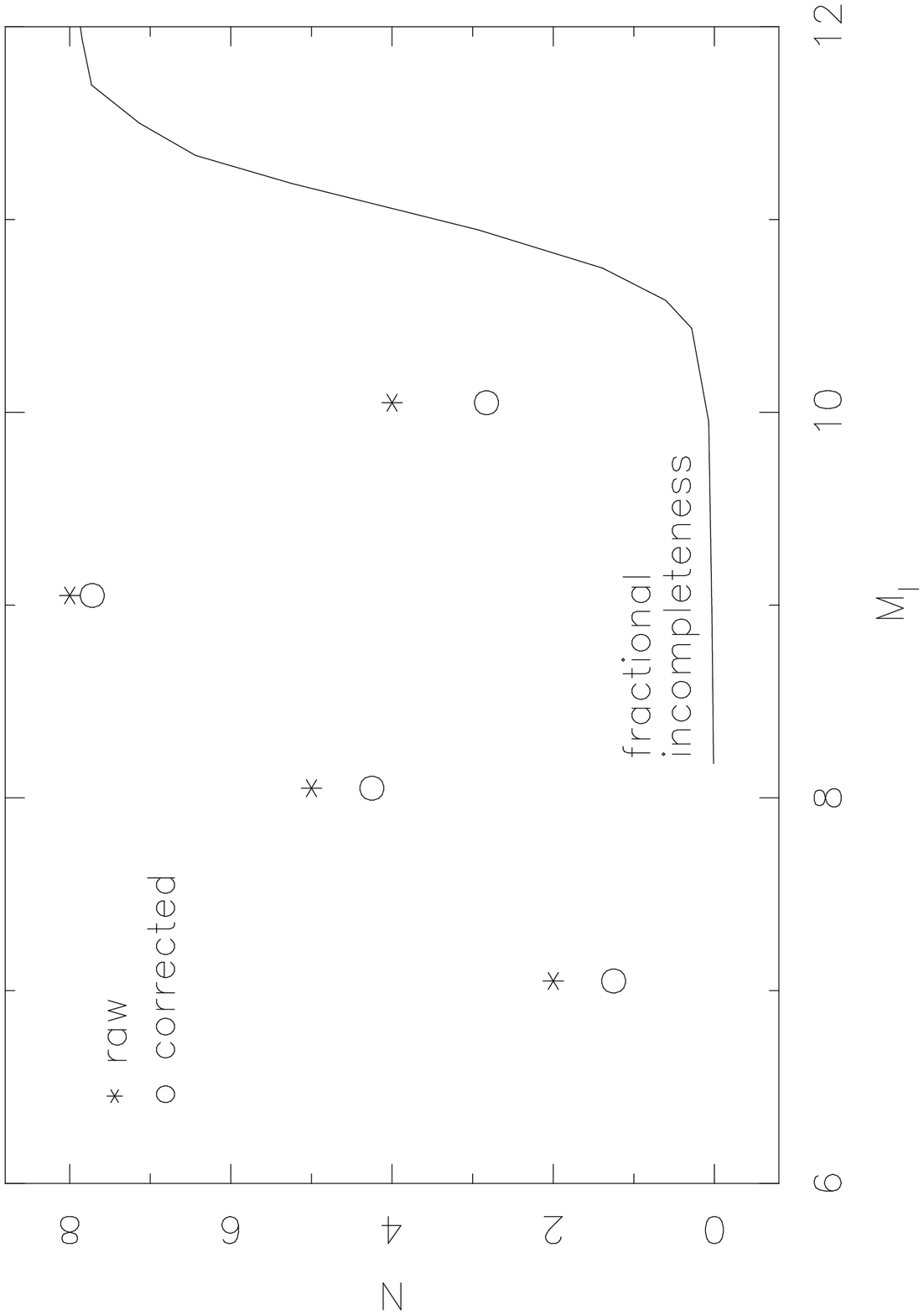,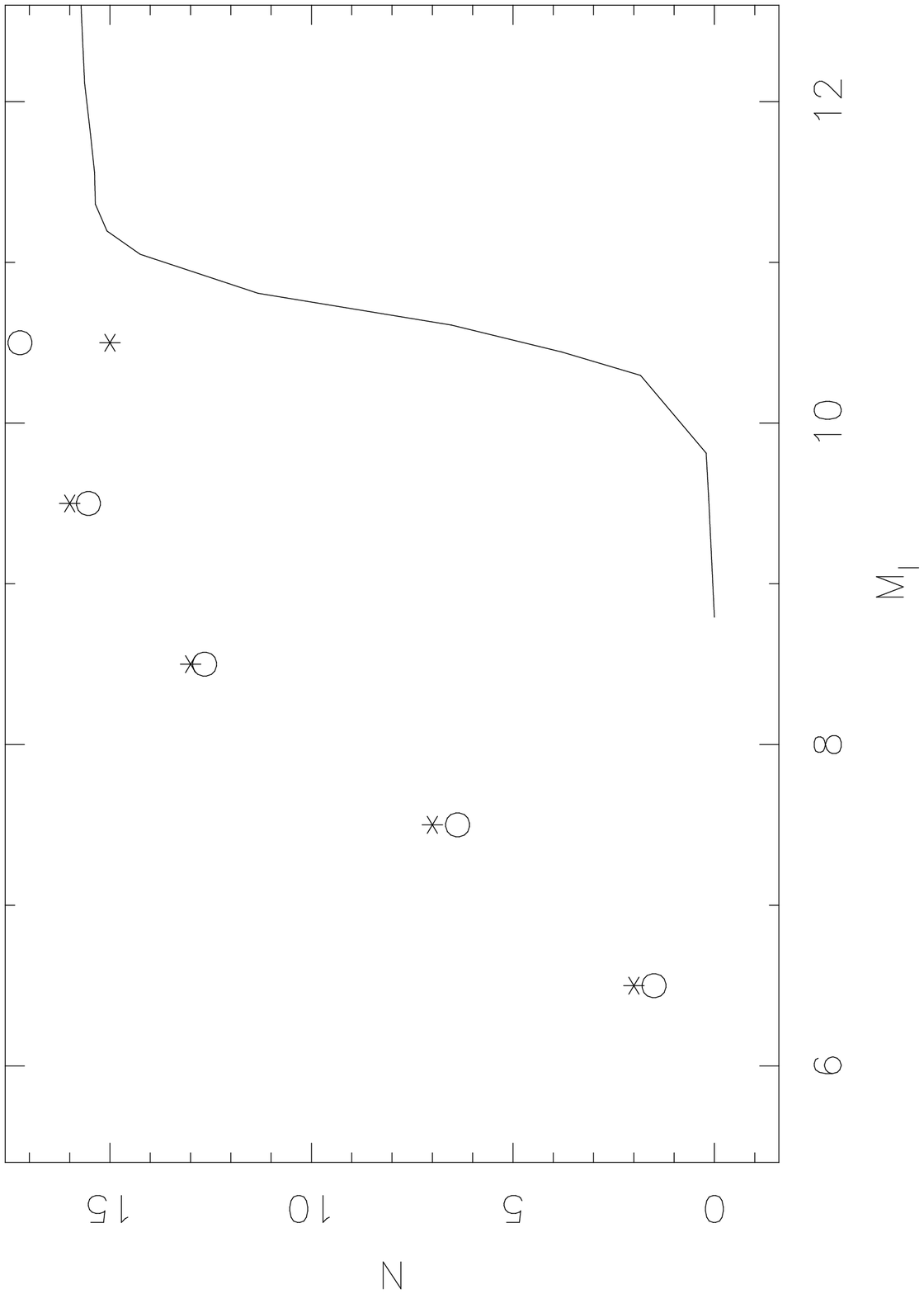]
{Main sequence luminosity function for (a) NGC 2420 and (b) NGC 2477.  The
star symbols are the raw LF and the open symbols are the LF corrected for
contaminating stars and incompleteness.  The incompleteness function for
each field is plotted as a solid line rising from 0\% to 100\%.}

\figcaption[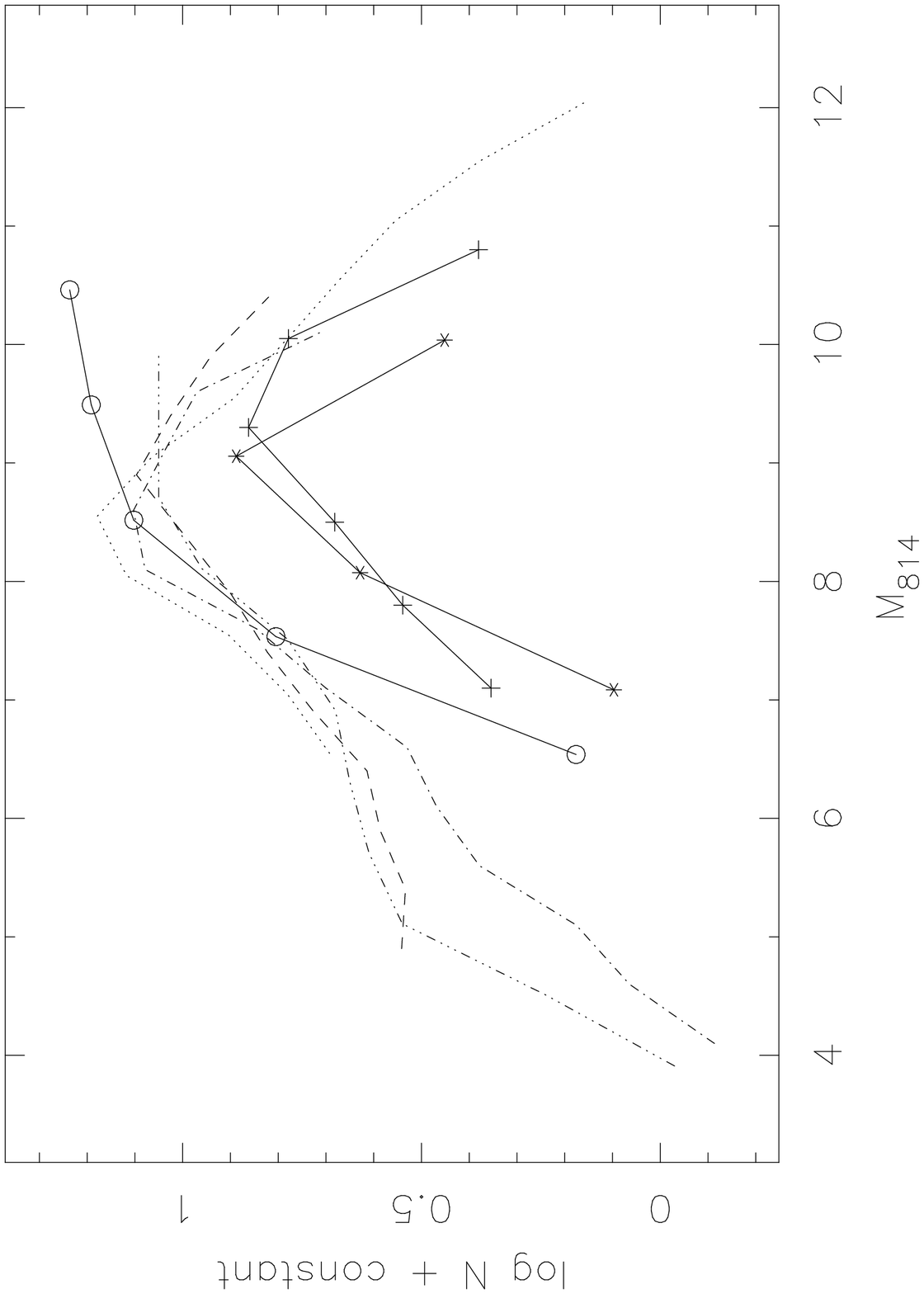]
{Globular and open cluster luminosity functions.  The solid lines are for
NGC 2477 (top, with ``o'' symbols), the solar neighborhood (middle, with
``+'' symbols), and NGC 2420 (bottom, with ``*'' symbols).  The dotted
line represents NGC 6397, the dashed line 47 Tuc, the dash single-dotted
line M15, and the dash double-dotted line $\omega$ Cen.  The horizontal
axis is $M_{814}$, the HST I-band absolute magnitude, which is very nearly
the standard Cousins I-band magnitude.  The numbers on the vertical axis
are correct for the open clusters, but the globular clusters have been
shifted 1 to 2 units downwards for ease of plotting.}


\figcaption[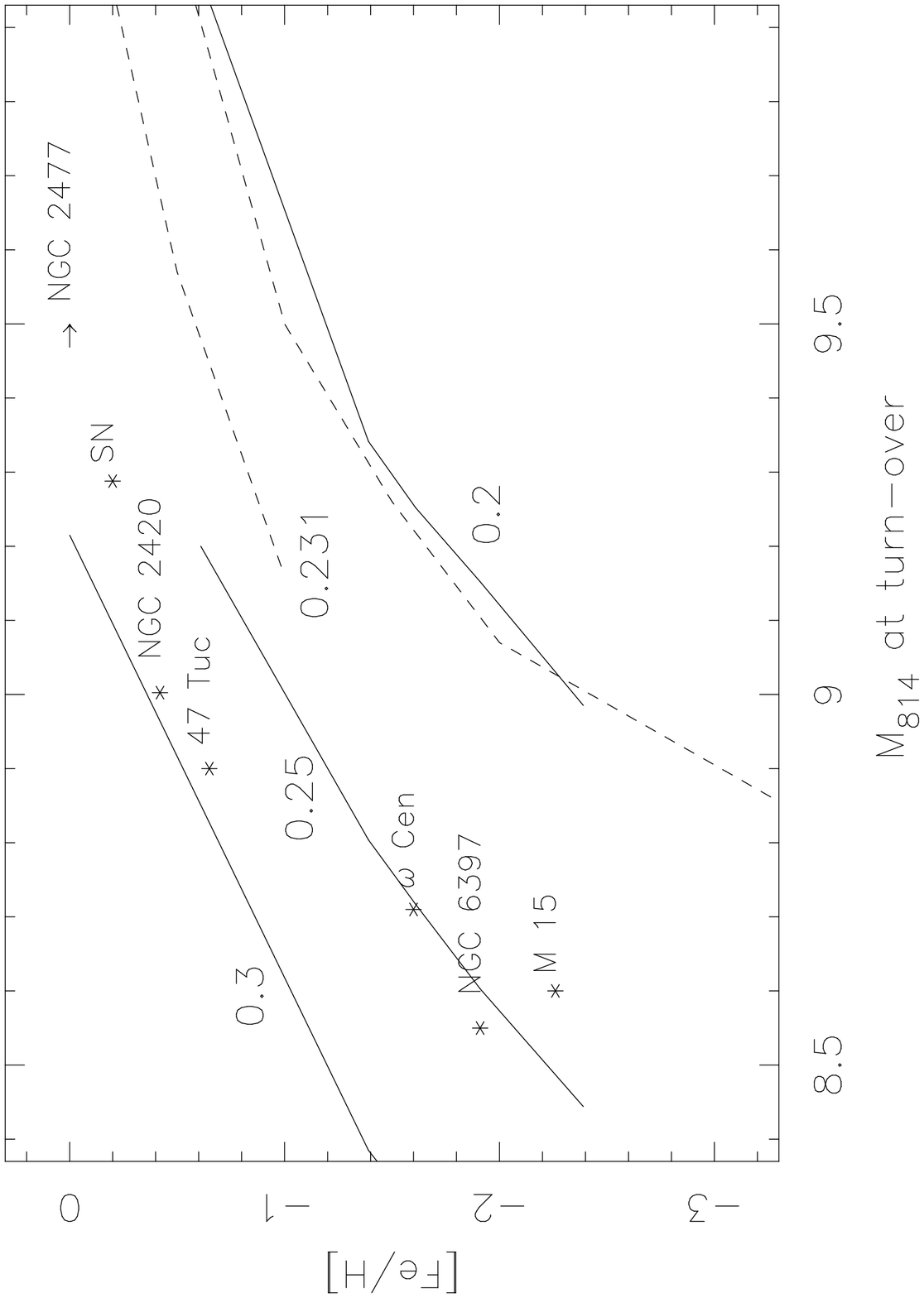]
{Absolute magnitude in F814W of the turn-over in the luminosity functions
as a function of metallicity.  The clusters are labeled.  The solar
neighborhood is indicated ``SN''.  Also plotted are labeled loci of
constant mass from Saumon (private communication) and Alexander {\it et
al.}\ (1996).  The theoretical loci for the Saumon models are actually in
$M_I$, but the difference between $M_I$ and $M_{814}$ is small.}

\end{document}